\documentstyle[preprint,eqsecnum,aps,prb]{revtex}

\begin{document}
\title{Anisotropic pinned/biased magnetization\\
in $SrRuO_3/SrMnO_3$ superlattices}
\author{P. Padhan and W. Prellier\thanks{%
prellier@ensicaen.fr}}
\address{Laboratoire CRISMAT, CNRS\ UMR 6508, ENSICAEN,\\
6 Bd du Mar\'{e}chal Juin, F-14050 Caen Cedex, FRANCE.}
\date{\today}
\maketitle

\begin{abstract}
The exchange coupling at the interfaces of magnetic superlattices consisting
of ferromagnetic $SrRuO_3$ and antiferromagnetic $SrMnO_3$ grown on ($001$)
oriented $SrTiO_3$ is studied with in-plane and out-of-plane orientations,
with respect to the substrate plane, of the cooling magnetic field. The
magnetization of the in-plane, field cooled hysteresis loop is lower than
the corresponding in-plane zero-field-cooled hysteresis loop. The
out-of-plane field cooled hysteresis loop is shifted, from the origin, along
the graphical magnetization axis. We attribute this irreversible rotation of
the moment to the pinning/biasing of spin in the $SrRuO_3$ layer in the
vicinity of interfaces by the antiferromagnetic $SrMnO_3$ layer.

\newpage
\end{abstract}

The discoveries of novel properties of magnetic films, multilayers and
micro- or nanostructures are of high potential to science and technology.
The new magnetic phenomena are in the field of nanomagnetism and spin
electronics\cite{1} based on multilayers composed of ferromagnetic ($FM$),
antiferromagnetic ($AFM$) and non-magnetic, either metallic or insulating,
materials. However, the $FM/AFM$ multilayers based on 3d - transition metal
compounds exhibit overdamped oscillatory exchange coupling\cite{2,3,4} and
unidirectional magnetic anisotropy\cite{5,6}. These two magnetic effects
originate from the coupling between the $FM$ layers, through the $AFM$
layer, and exchange coupling at the $FM-AFM$ interfaces. The unidirectional
magnetic anisotropy is commonly characterized by an exchange field, the
field through which the center of the hysteresis loop of the ferromagnet
shifts from zero. This phenomenon is generally known as exchange bias. It is
usually observed upon cooling the $FM/AFM$ system through the N\'{e}el
temperature $T_N$ of the $AFM$, to a temperature below the Curie temperature
of the FM in the presence of a magnetic field. The atomic and magnetic
structures at the $AFM/FM$ interfaces play a decisive role in the
interaction mechanism and thus also for the magnetic properties of the
exchange coupled system. The lattice mismatch and strain relaxation can also
lead to crystallographic, and/or magnetic reconstructions and relaxations at
the interface, which will influence the exchange coupling behavior. The
exchange bias phenomenon has been observed in a wide variety of $FM$ and $%
AFM $ systems including simple spin structures at the interface to the FM
layer, such as polycrystalline layers or materials that do not have
uncompensated planes of spins in any directions. In general, exchange bias
is established through field cooling in-the-film-plane where the magnetic
easy axis of soft ferromagnetic materials normally lies in the plane.
Recently, Maat {\it et al.}\cite{7} have shown that exchange bias can also
be observed for magnetization perpendicular to the film plane in $Co/Pt$
multilayers biased by $CoO$. They investigated the biasing in various
directions and found it to be substantially larger within the sample plane,
which they related to the anisotropy of the single-q spin structure of the $%
CoO$.

Most of the previous studies have been performed on metallic compounds, but
we have decided to study similar phenomena on transition metal oxide
multilayers, grown by laser ablation.\ Thus, we have synthesized
superlattices consisting of ferromagnetic $SrRuO_3$ ($SRO$) and
antiferromagnetic $SrMnO_3$ ($SMO$) layers, and we report on the structural
and magnetic properties of this superlattice system. This magnetic structure
exhibits pinned/biased moment below a certain critical field ($H_P$ $\approx 
$ $2$ $tesla$), and the effects of magnetic fields below $H_P$ are presented
in this article. The pinning effect can be realized in the field-cooled (FC)
hysteresis loop. We found that orientation of the cooling magnetic field
along the film plane quenches magnetic moments while an out-of-plane cooling
field shifts the hysteresis loop, from the origin, along the moment axis. We
attribute this irreversible rotation of the moments to the pinning/biasing
of spin in the $SRO$ in the vicinity of the interfaces by the
antiferromagnetic $SMO$.

A multitarget pulsed laser deposition system was used to prepare the thin
films and superlattices of $SRO$ and $SMO$ on ($001$)-oriented $SrTiO_3$ ($%
STO$) substrates (lattice parameter $a_{STO}=3.905\,\AA $). The thin films
and multilayers were deposited at $720$ $^{\circ }C$ in ambient oxygen at a
pressure of $30$ $mTorr$. The deposition rates (typically $\symbol{126}0.26$ 
$\AA /pulse$) of $SRO$ and $SMO$ were calibrated individually for each laser
pulse of energy density $\symbol{126}3$ $J/cm^2$. After the deposition, the
chamber was filled to $300$ $Torr$ of oxygen at a constant rate and then the
samples were cooled to room temperature at rate of $20$ $^{\circ }C/\min $.
The superlattice structures were synthesized by repeating the bilayer
comprising of $20$-($unit$ $cell$, $u.c.$) $SRO$ and $n$-($u.c.$) $SMO$, $15$
times, with $n$ taking integer values from $1$ to $20$. In all samples, the
bottom layer is $SRO$ and the modulation structure was covered with $20$ $%
u.c.$ $SRO$ to keep the structure of the topmost $SMO$ layer stable. The
periodic modulations in composition were calculated using established
deposition rates of $SRO$ and $SMO$ obtained from the positions of
superlattice reflections in $X$-ray $\theta -2\theta $ scans. The epitaxial
growth and structural characterization of the multilayer and single layer
films were performed using $X$-ray diffraction (XRD), electron dispersive
spectroscopy ($EDS$) and transmission electron microscopy ($TEM$)\cite
{strain}. The magnetization ($M$) measurements were performed using a
superconducting quantum interference device based magnetometer ($Quantum$ $%
Design$ $MPMS-5$). The magnetization measurements were carried out by
cooling the sample below room temperature in the presence/absence of
magnetic fields along the [$100$] and [$001$] directions of the $STO$
substrate. The orientation of the magnetic field during the field-cooled
measurements remains similar to that of the cooling field.

Fig. 1 shows the ($002$) reflections of the samples recorded during the $%
\theta -2\theta $ x-ray scans of the thin films of $SRO$ and $SMO$,
respectively, deposited on ($001$)-oriented $STO$. The diffraction profiles
show only ($00l$) reflections from both the film and substrate, indicating
the epitaxial growth of $SRO$ and $SMO$ on (001)-oriented $STO$. The XRD\
graph of a $200$ $u.c.$ thick film of $SRO$ is shown in the upper panel
(Fig.1a) whereas the lower panel (Fig.1b) shows the scan of a $40$ $u.c$.
thick film of $SMO$ covered with a $10$ $u.c.$ thick $STO$ (this $10$ $u.c$.
layer of $STO$\ is only present in order to protect the $SMO$ layer). The
lattice parameter of bulk $SRO$ ($a_{SRO}$ $=$ $3.93$ $\,\AA $) is larger
than $a_{STO}$, with a lattice mismatch of $+$ $0.6$ $\%$, whereas the
lattice parameter of $SMO$ ($a_{SMO}$ $=$ $3.805$ $\,\AA $) is smaller than $%
a_{STO}$, with a lattice mismatch of $-$ $2.5$ $\%$. The nature of
substrate-induced stress for the epitaxial growth of $SRO$ and $SMO$ on $STO$
is the opposite. In our film, this epitaxial correlation is confirmed from
the angular positions of $SRO$, $SMO$ and $STO$ (see Fig. 1).

The $\theta -2\theta $ x-ray scans of the superlattices consisting of $SRO$
and $SMO$ deposited on ($001$)-$STO$ also show the ($00l$) diffraction peaks
of both constituents and the substrate, confirming the $c$-axis orientation
with a pseudocubic lattice parameter. In Fig. 2 we show the diffracted x-ray
intensity with the $2\theta $ range around the ($001$) reflection of these
pseudocubic perovskite superlattices of ($20.u.c.$)$SRO/$($1.u.$ $SMO$). The
presence of second order satellite peaks on either side of the fundamental ($%
001$) reflection clearly shows periodic chemical modulation of the
constituents. For the quantitative refinement of this sample, we have used
the $DIFFaX$ program\cite{diffax} to simulate the diffraction intensity for
various diffraction angles. The simulated diffraction profiles of this
sample close to the ($001$) reflection is shown in Fig. 2. The positions and
relative intensity ratios of the satellite peaks in the measured $\theta
-2\theta $ x-ray scan are in good agreement with the simulated profile. The
values of the superlattice periods, estimated\cite{ref} from the angular
position of the satellite peaks in the $\theta -2\theta $ x-ray scans, are
also in agreement with the calculated value\cite{strain}.

$SrRuO_3$\ is known as a metallic ferromagnet, with a Curie temperature ($%
T_C $) $\sim 160$ $K$ in its bulk form\cite{sro}. The magnetic properties of
a relaxed $\symbol{126}$ $200$ $u.c.$ thick $SRO$ film deposited on $STO$ is
shown in Fig. 3. The temperature-dependent magnetization of the film is
similar to that of its bulk, while its zero-field-cooled ($ZFC$) magnetic
hysteresis at $10$ $K$ shows a magnetically easy axis along the [$001$]
direction of the substrate. In the hysteresis loop the saturation
magnetization ($M_S$), saturation field ($H_S$) and coercive field ($H_C$),
measured along the easy axis of this film, are $1.46$ $\mu _B/Ru$, $0.4$ $%
tesla$ and $0.17$ $tesla$, respectively. For the hysteresis loop along the
easy axis, the negligibly small difference between the remanent
magnetization ($M_R$) and $M_S$ indicates a coherent rotation of
magnetization in a single domain film. As the sample is cooled below room
temperature down to $10$ $K$, in the presence of a $0.1$ $tesla$ magnetic
field, the shape of the hysteresis loop is similar to its $ZFC$ hysteresis
loop. In contrast, $SrMnO_3$ is an antiferromagnet\ with Neel temperature ($%
T_N$) close to $260$ $K$\cite{smo}, which crystallizes into a cubic
structure (Fig. 1b) when sandwiched between perovskite layers inside a
superlattice\cite{strain,9}.

The low field ($0.01$ $tesla$) temperature dependent $ZFC$ and $FC$
magnetizations of the sample with $n=1$ are shown in Fig. 4(a). The $ZFC$
magnetization with in-plane magnetic field increases slowly when heated
above $10$ $K$, reaching a maximum at $\sim $ $150$ $K$ and then decreasing
slowly upon further heating to room temperature. The $FC$ magnetization,
with a $0.01$ $tesla$ cooling field, remains the same when heated from 10 K
up to $\sim $ $50$ $K$. Upon further heating, the magnetization decreases
slowly and than rapidly in the temperature range of $100$ $K$ to $160$ $K$.
Above $160$ $K$ the value of magnetization is close to that of its $ZFC$
value. The large difference in the $ZFC$ and $FC$ magnetizations below $160$ 
$K$ suggests freezing of the moments of $SRO$ by the $SMO$ in the $ZFC$
state. The low field temperature dependent magnetization along [$100$] and [$%
001$] directions of the substrate for the sample with $n$ $=$ $4$ is shown
in Fig. 4(b). In this magnetic structure, the alternate stacking of $SRO$
and $SMO$ leads to the magnetic inhomogeneity along the [$001$] direction of
the substrate, which may lead to the low field anisotropy in the
magnetization along the [$100$] and [$001$] directions of the substrate.
Another possible source of the low field anisotropy could be due to the
observed anisotropy in the SRO layer (Fig. 3b). To establish the origin of
these two effects we have measured the field dependent ZFC and FC
magnetizations of these samples at $10$ $K$. The hysteresis loops are shown
in Fig. 5. The measurements are based on two aspects: firstly by the
response of the magnetic anisotropy of the sample to the direction of
magnetic field, and secondly by the effect of a cooling field on the
magnetic configuration of the sample.

The $ZFC$ hysteresis loop measured in the field range of $\pm 1$ $tesla$
(larger than the saturation field of $SRO$) shows lower values of $M_R$ and $%
H_C$ compared with the thin film of $SRO$ (Fig. 3b). The magnetization
increases gradually as the magnetic field increases and the hysteresis loop
does not reveal a distinct $H_S$ and $M_S$. The $ZFC$ hysteresis loops with
a magnetic field oriented along the [$100$] and [$001$] directions of the
substrate for the sample with $n=2$ are shown in Fig. 5(a). The loop shape
indicates that the effect of $AFM$ $SMO$ on the magnetic configurations of $%
SRO$ is stronger for a field perpendicular to the film plane. The in-plane
hysteresis loop is symmetric with respect to the magnetization as well as
the field axis, while the out-of-plane hysteresis loop is symmetric along
the field axis, with a negligibly small shift along the magnetization axis.
This shift of the hysteresis loop from the origin along the magnetization
axis is due to the presence of a small magnetic field remaining from when
the sample was cooled below room temperature. This intrinsic effect is
clearer when the sample is cooled to $10$ $K$ below room temperature in the
presence of a magnetic field.

The in-plane $ZFC$ and $FC$ hysteresis loops for the sample with $n=3$ are
shown in Fig. 5(b). These hysteresis loops are symmetric with respect to the
field and the magnetization axis (graphical axis).The magnetization of the
in-plane $ZFC$ hysteresis loop decreses as the sample is cooled below room
temperature in the presence of $0.1$ $tesla$ magnetic field. However, a
similar effect is not observed for the out-of-plane $ZFC$ and $FC$
hysteresis loops. On cooling below room temperature down to $10$ $K$ in the
presence of a $0.1$ $tesla$ magnetic field the out-of-plane hysteresis loop
at the SRO/SMO superlattices shifts, from the origin, along the
magnetization axis. The out-of-plane $ZFC$ and $FC$ hysteresis loops of the
superlattice with $n=5$ and $10$ are shown in Fig. 5(c) and 5(d),
respectively. The values of $H_C$ are equal and opposite for both the
increasing and decreasing branches of the $FC$ loops whereas the values of $%
M_R$ are different with the same sign. The magnetization of the $FC$ loop
increases gradually as the magnetic field increases and neither reveal a
distinct $H_S$ nor $M_S$.

The magnetic interaction across the interface between an $FM$ spin system
and an $AFM$ spin system is known as exchange coupling. This interfacial
magnetic coupling depends strongly on the spin configuration at the $FM-AFM$
interfaces, which occurs due to the crystallographic, and/or magnetic
reconstructions, and relaxations at these interfaces. In SRO/SMO
superlattices we have observed the variation of relaxation at the SRO-SMO
interfaces with the SMO layer thickness\cite{strain}. Thus the magnetic
coupling and the spin configuration at these interfaces of $FM$ $SRO$ and $G$%
-type $AFM$ $SMO$, depend on the $SMO$ layer thickness. When the $SMO$ layer
thickness is 1 u.c., the interfacial spin configuration is associated with
the $3D$-coordination of $RuO_6$ and $MnO_6$ and the in-plane staggered
pattern spin arrangement in $SMO$. This is the source of spin frustration at
the $SRO/SMO$ interfaces as well as the spin canting in the $SRO$ layer in
the vicinity of these interfaces. However, as the $SMO$ layer thickness
increases above 1 u.c. the interface spin arrangement is influenced by
another component due to the staggered pattern spin in $SMO$ along the ($00l$%
) planes. In otherwords, above 1 u.c. the interface spin arrangement is
influenced by the in-plane and out-of-plane staggered patterns. The increase
of the ($00l$) planes in $SMO$ increases spin canting in the $SRO$ and
saturate for the higher values of $SMO$ layer thickness. A similar effect is
observed for the magnetic moment of the superlattice for various $SMO$ layer
thicknesses.

Since the $T_C$ of $SRO$ is smaller than the $T_N$ of $SMO$ when cooling the
superlattices below room temperature, the exchange coupling at the
interfaces will vary at different temperature zones. At the interfaces the
exchange between the transition metal ions, when both the materials are
paramagnetic ($PM$) at room temperature, changes as SMO becomes AFM (at $T_C$
$<$ $T$ $<$ $T_N$) and SRO becomes FM (at $T<$ $T_C$). At $T<$ $T_C$ the
coupling energy at the interfaces between $SRO$ and $SMO$ layers is
proportional to the magnetic field\cite{bean}. Thus, the spin configurations
of the $SRO/SMO$ superlattices are influenced by the magnetic field, its
orientation and the thermal energy during the $FC$ state. Therefore, a
difference can be expected in the coupling energy at the interfaces in the $%
FC$ and $ZFC$ state. However, in this superlattice the strength and nature
of coupling depend on the orientation of the magnetic field. Assuming that
the bulk spin configuration is preserved\cite{5}, we have considered that
the $SMO$ is rigid with respect to the orientation at low field ($<2$ $tesla$%
) and the moments lie in-the-plane of the film. According to this assumption
the spin configuration of the superlattices are in-plane when the magnetic
field is along the [$100$] direction while the spins of $SRO$ and $SMO$ are
oriented perpendicular to each other for the magnetic field along the [$001$%
] direction of the substrate. Since the anisotropy axis of $SMO$ is fixed,
the magnetic field along the easy axis of $SMO$ decreases the angle between
the magnetization of $SRO$ and the easy axis of $SMO$ while their angular
separation increases as the magnetic field rotates by $90{{}^{\circ }}$. For
both orientations, the magnetic field leads to different Zeeman energies,
and this could be the source anisotropy (quenching of the in-plane moment
and shifting of the out-of-plane hysteresis loop from the origin along the
moment axis). The spins close to the interfaces in $SRO$ are canted\cite
{cant1}, and their rotations are reversible as seen in the $ZFC$ hysteresis
loop, but when the cooling field is applied, the coupling energy increases
leading to an irreversible rotation below the critical magnetic field.
\smallskip We attribute this irreversible rotation of the moment to the
pinning/biasing of spin in the $SRO$ in the vicinity of interfaces by the $%
AFM$ $SMO$.

In conclusion, we have observed pinned/biased moment in the $FM/AFM$
superlattices consisting of a ferromagnetic $SRO$ and antiferromagnetic $SMO$
bilayer. The magnetization of the $FM$ $SRO$ layer is suppressed in the $%
SRO/SMO$ superlattices. We attribute this to the randomly pinned/biased
moments in the $SRO$ layer in the vicinity of the interfaces due to the
strong exchange coupling at the $SRO$-$SMO$ interfaces. These superlatices
show a shift of the out-of-plane $ZFC$\ hysteresis loop, from the origin,
along the magnetization axis, wheras the magnetc moment quenches the
in-plane hysteresis loop. This anisotropic effect can be viewed as the
oriented ($FM$ or $AFM$) pinning/biasing of moments in the $SRO$ in the
vicinity of the interfaces caused by the cooling field. The exchange
coupling between the $SRO$ and $SMO$, due to the modified $3D$-coordination
of $Mn$ and $Ru$ ions and the anisotropic nature of $SMO$, are responsible
for anisotropic pinned/biased moments with the orientations of the magnetic
field. As our understanding of the phenomena at the interfaces of magnetic
multilayers is growing, it is hoped that these results may bring new
insights about such important issues.

Acknowledgments:

We thank Prof.\ B.\ Mercey for the helpful discussions. We also thank Dr.\
H.\ Eng for his careful reading of this article. We greatly acknowledge the
financial support of the Centre Franco-Indien pour la Promotion de la
Recherche Avancee/Indo-French Centre for the Promotion of Advance Research
(CEFIPRA/IFCPAR) under Project N${{}^{\circ }}$2808-1.

\bigskip

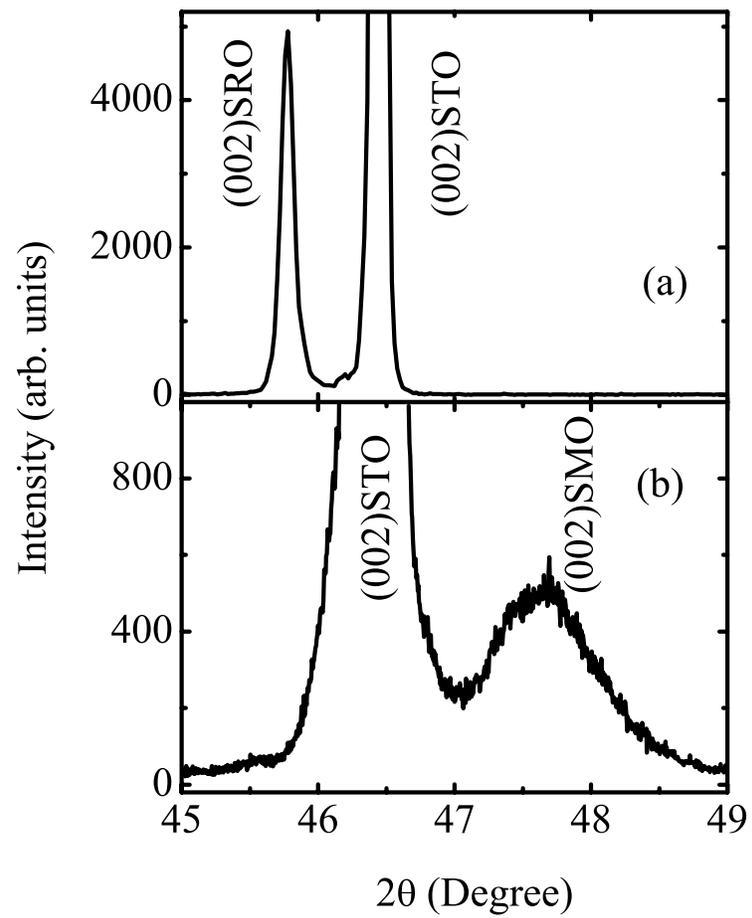
Fig. 1: X-ray diffraction profiles of the samples recorded around the $002$
reflection of $STO$. (a) $200$ $u.c.$ $SRO$ on $STO$ and (b) bilayer of ($10$
$u.c.)STO/(40$ $u.c.)SMO$ on $STO$.

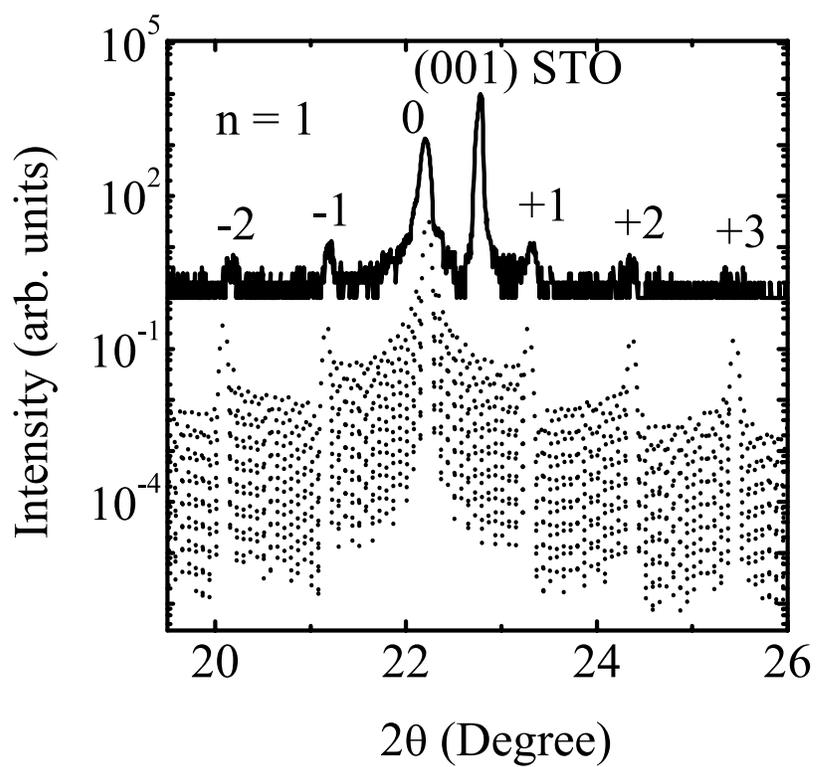
Fig. 2: Measured and simulated $\Theta -2\Theta $ spectra for ($20$ $u.c.$)$%
SRO$/($1$ $u.c.$) $SMO$ superlattice around the ($001$) reflection of $STO$.

Fig. 3(a) $FC$ magnetization of the ($200$ $u.c.$)$SRO$ on $STO$ at
different temperatures with a $0.1$ $tesla$ in-plane magnetic field. (b)
Isothermal ($10$ $K$) magnetization ($ZFC$\ and $FC$) of the ($200$ $u.c.$)$%
SRO$ on $STO$ with various fields oriented along the [$100$] and [$001$]
directions of the $STO$ substrate.

Fig. 4(a) $ZFC$ and $FC$ magnetization of the superlattice with $n$ $=$ $1$
at different temperatures with a $0.1$ $tesla$ in-plane magnetic field. (b) $%
FC$ magnetization of the superlattice with $n$ $=$ $4$ at different
temperature at $0.1$ $tesla$ magnetic field oriented along the [$100$] and [$%
001$] directions of the $STO$ substrate.

Fig. 5(a) $ZFC$ magnetization of the superlattice with $n$ $=$ $2$ at $10$ $K
$ with various magnetic fields oriented along the [$100$] and [$001$]
directions of the $STO$ substrate. (b) $ZFC$ and $FC$ magnetization of the
superlattice with $n$ $=$ $3$ at $10$ $K$ with various in-plane magnetic
fields. (c) and (d) $ZFC$ and $FC$ magnetization of the superlattice with $n$
$=$ $5$ and $10$, respectively, at $10$ $K$ with various out-of-plane
magnetic fields.

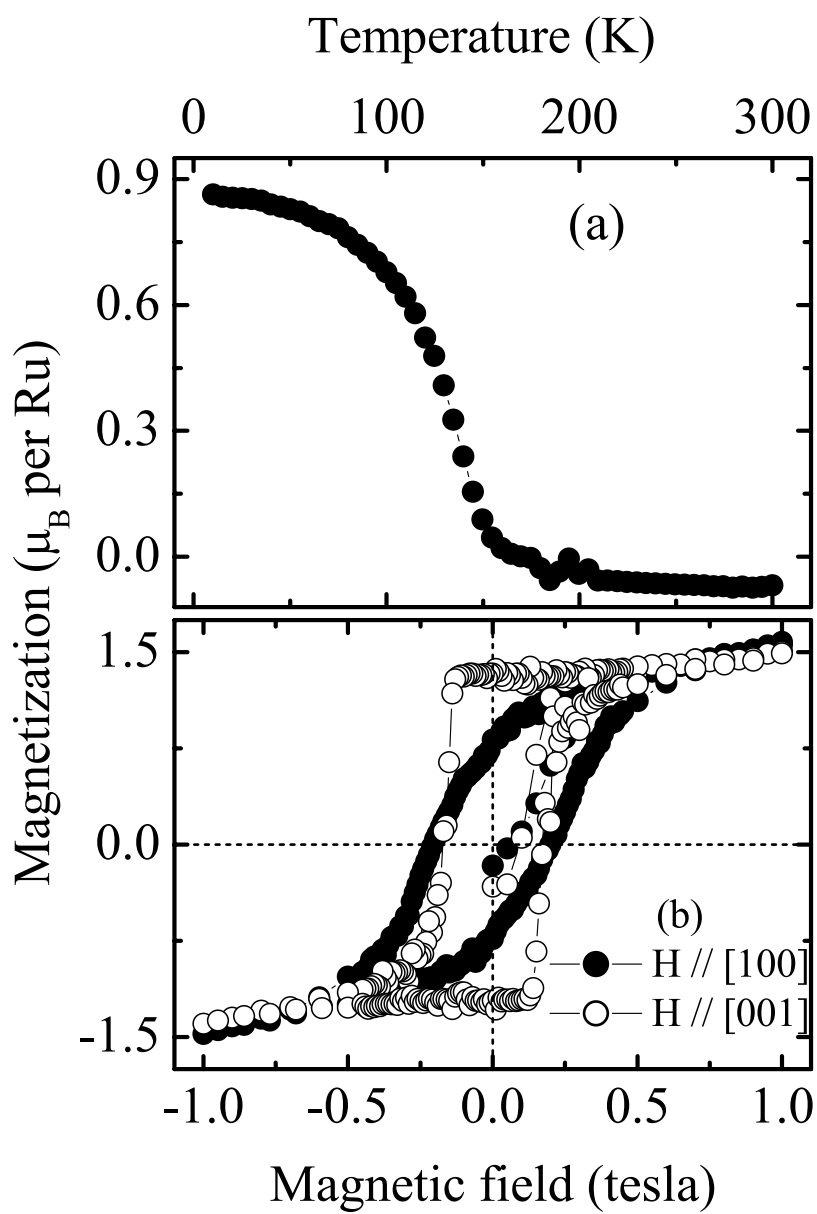
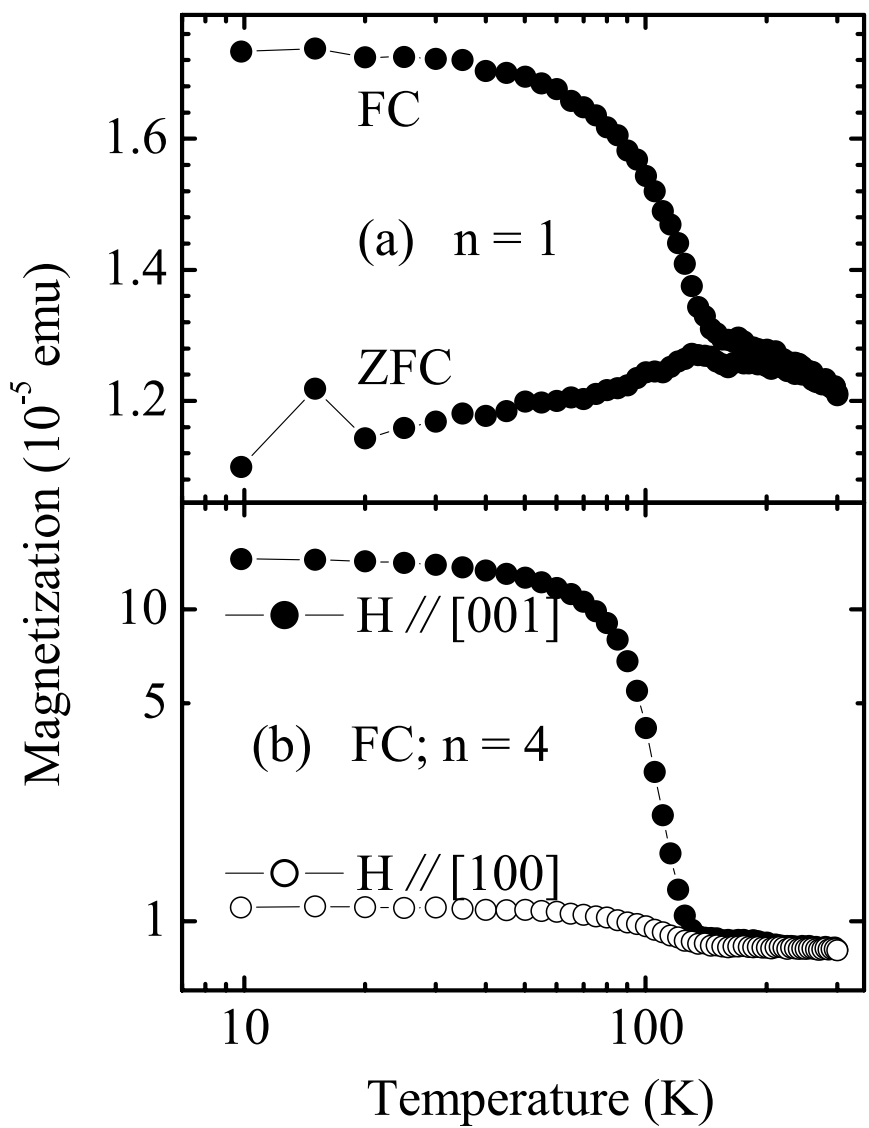
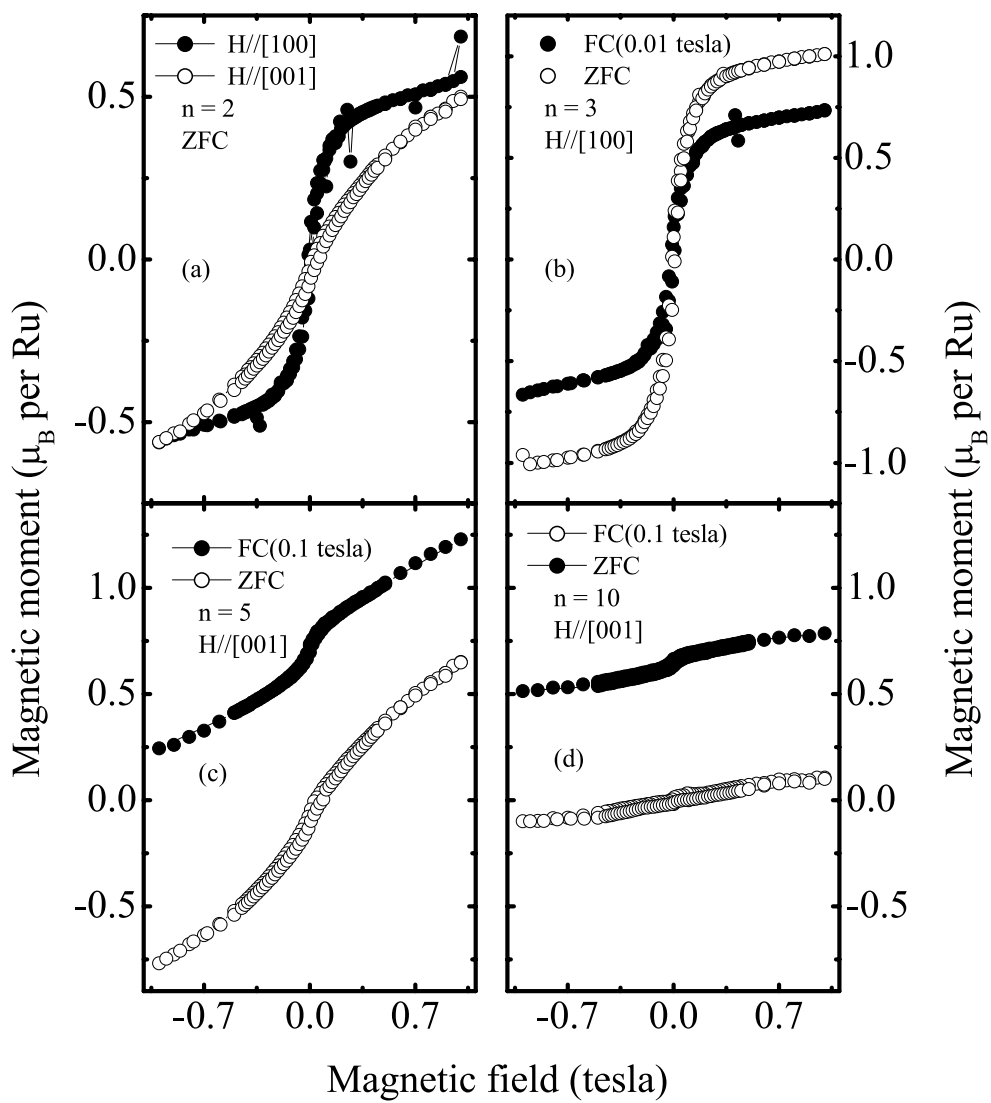
\end{document}